\documentclass{article}
\usepackage[utf8]{inputenc}

\title{Towards a Soft Faceted Browsing Scheme for Information Access}
\author{Yinan Zhang\thanks{University of Illinois at Urbana-Champaign; email: yzhng103@illinois.edu}, Parikshit Sondhi\thanks{WalmartLabs; email: psondhi@walmartlabs.com}, Anjan Goswami\thanks{WalmartLabs; email: agoswami@walmartlabs.com}, ChengXiang Zhai\thanks{University of Illinois at Urbana-Champaign; email: czhai@illinois.edu}}
\date{}

\usepackage{hyperref}
\usepackage{empheq}
\usepackage{bm}
\usepackage{dsfont}

\usepackage{amsthm}

\theoremstyle{definition}
\newtheorem*{definition}{Definition}

\begin{document}
\maketitle

\begin{abstract}
Faceted browsing is a commonly supported feature of user interfaces for access to information.
Existing interfaces generally treat facet values selected by a user as hard filters and respond to the user by only displaying information items strictly satisfying the filters and in their original ranking order.
We propose a novel alternative strategy for faceted browsing, called soft faceted browsing, where the system also includes some possibly relevant items outside the selected filter in a non-intrusive way and re-ranks the items to better satisfy the user's information need.
Such a soft faceted browsing strategy can be beneficial when the user does not have a very confident and strict preference for the selected facet values, and is especially appropriate for applications such as e-commerce search where the user would like to explore a larger space before finalizing a purchasing decision.
We propose a probabilistic framework for modeling and solving the soft faceted browsing problem, and apply the framework to study the case of facet filter selection in e-commerce search engines.
Preliminary experiment results demonstrate the soft faceted browsing scheme is better than the traditional faceted browsing scheme in terms of its efficiency in helping users navigate in the information space.
\end{abstract}

\section{Introduction}
Faceted browsing \cite{hearst:2009} is a commonly supported feature of user interfaces for access to information.
It allows a user to flexibly drill down to a specific set of information items as needed.
Existing interfaces generally treat facet values selected by a user as hard filters and respond to the user by only displaying information items strictly satisfying the filters and in their original ranking order. 
Such an interface is easy to interpret and the user has full control of what to be shown. 

However, the relevant information items that users are interested in often do not satisfy the filters they select.
% This happens for the following two reasons.
% First, the facet values or filters with ranges are not personalized and thus there may not exist a filter that completely matches what the user is interested in.
% For example, a price filter may use ranges such as ``below \$200'', ``[\$200, \$500]'', etc, but a user's preference range might be ``[\$100,\$300]".
% Second, a user may not know {\em in advance} the optimal facet values for the items interesting to them, which is often the case in exploratory search, thus their chosen filters again may not match accurately with their preferences. 
This happens usually because a user may not know {\em in advance} the optimal facet values for the items interesting to them, which is often the case in exploratory search, thus their chosen filters may not match accurately with their preferences.
Indeed, as we will discuss later in the paper, our analysis of the search log of a commercial e-commerce search engine \cite{walmart}
% \footnote{http://www.walmart.com/}
suggests that around 43\% of users' selected price filters do \emph{not} cover the price of their eventually purchased product.
The problem of mismatching between users' selected facet filters and their information seeking goals is very frequently observed in other types of facet filters in e-commerce search engines, e.g. brand, color, customer rating, etc., as well as information access systems in different domains, such as search engines for online library catalogs.
In all these cases, since users cannot find any item interesting to them after the facet filter is applied, they have to de-select the filter and then either scan the unfiltered list or try other facet selections, which greatly affects the navigational efficiency as well as the overall user experience.
These problems are caused by the fact that users often have a knowledge gap about the information items to be found, and thus their actions are often probabilistically rather than deterministically related to their goals in human-computer interactions.

Facing such probabilistic human actions, a traditional faceted browsing system always responds to human actions conservatively, and humans totally rely on themselves to resolve the issue resulted from their probabilistic nature, e.g. via means of reverting or modifying the filters.
This approach has the apparent advantage on the system's side that the system is always highly predictable, and it is in many cases the best the system could do in a world with little knowledge about user actions.
However, in the contemporary era of massive user log data accumulation and fast advancement in machine learning methodologies, the na{\"i}ve ``pretending-to-be-blind'' strategy of the traditional faceted browsing scheme becomes insufficient, calling the need for more intelligent approaches capable of sharing with humans the burden of resolving their probabilistic-flavored actions without sacrificing much of the system's predictability.

In this work, we propose a novel ``soft faceted browsing'' scheme where we not only show items that match the user's selected facet filter, but also show a few possibly relevant items that do not match the selected filter in a non-intrusive way, and re-rank the items to best satisfy the user's information need.
% and we establish a principled Bayesian probabilistic framework for achieving the ``soft faceted browsing'' scheme.
Such a ``soft facet filter'' interface has several potential benefits: (a) the additional items shown may be relevant to the user's interest, and thus save the user effort on selecting another filter; (b) it makes a user aware of existence of possibly relevant items that do not match the selected filter; and (c) it enables the system to learn about a user's interest more effectively by ``testing'' items not matching the selected filter.
% To enable the user to easily identify the original items satisfying the selected filter, the inserted new items will be displayed differently (e.g., indented form or distinct color) to signal that they are outside the selected attribute values.
% Figure 1 illustrates a soft faceted browsing interface in comparison with a traditional interface, where the left figure shows the top portion of a ranked list of items returned by a commercial e-commerce search engine \cite{walmart} as a response to the query ``luggage'', and the middle and right figures show the items returned by the system after the user selects the ``[\$20, \$50]'' filter as determined by the traditional ``hard'' faceted browsing scheme and our proposed soft faceted browsing scheme, respectively.
% We see that two additional items were inserted into the list that are outside the user's selected price range, and they turn out to be decent choices for the user though they're priced slightly above \$50.
% In this prototype system, we employed ``recommended'' tags to highlight the out-of-filter items, but there could potentially be many alternative strategies, e.g. using different color, indentation, using smaller fonts, etc.
% As an initial step towards the soft faceted browsing scheme, however, we mainly focus on demonstrating its advantage over the traditional ``hard'' faceted browsing strategy and establishing a principled theoretical framework for realizing such a new browsing strategy in this study, and consider the interface design work as an orthogonal research direction that we leave to future studies.

To implement the new interface, the main technical challenge is to select the ``right" extra items to add to the list and to optimize their ranking. To this end, we propose a principled Bayesian probabilistic framework to characterize a user's preferences over different items and their facet selection actions, and we infer users' goals given their selected facet filters based on the Bayesian framework.
The parameters in the user action model can be learned via another separate layer of Bayesian inference computation from past user search log.
We apply our proposed framework to study some representative facets in e-commerce search engines to demonstrate the generality and effectiveness of the framework.
%We show that compared to the traditional faceted browsing scheme, our soft faceted browsing scheme could better %characterize user actions in reality and thus enables smarter interactions between users and the system.
Preliminary experiment results from e-commerce search engines reveal that the soft faceted browsing scheme is superior to the traditional faceted browsing scheme in terms of its efficiency in helping users navigate in the information space.

\section{Related Work}

A large body of research work in the field of Human-Computer Interaction (HCI) has been focused on the area of faceted browsing \cite{hearst:2009}, where users interact with an information access system via means of drilling down to subsets of the information space by selecting facet values.
% , and it has been shown to be an effective method for the users to efficiently and effectively navigate to their interested information item in many domains, e.g. image search \cite{van:2010:faceted}.
While most of the work addresses either the interface design, e.g. \cite{yee:2003:faceted}, or the optimal selection of facet values to be displayed to the user, e.g. \cite{kashyap:2010:facetor}, little attention has been paid to the \emph{probabilistic} nature of the users' facet filter selection actions and how the system optimally reacts to such actions.
We introduce formal models to address such probabilistic nature of user actions for the first time, and propose an alternative and more effective way for the system to optimally respond to user actions.

% Our work also broadly shares certain characteristics with a few other information retrieval research areas, and one example is search result interleaving.
The model we propose guides the system to inject certain ``out-of-filter'' items into the list of ``within-filter'' items, and in such a sense our work resembles some recent information retrieval research work in the area of search result interleaving \cite{chuklin:2013:evaluating,radlinski:2013:optimized}.
However, these studies mainly focused on interleaving search results for the purpose of comparing multiple search systems/methods, while we are aiming at optimally satisfying users' information need without considering any comparison across the interleaved sets of results.

Our framework takes into account both the original ranking of the items and how well each item satisfies the user's selected facet filter when coming up with an optimal ranking for the user's information need, and such an approach shares certain common characteristics with the information retrieval research work on rank aggregation \cite{dwork:2001:rank,liu:2007:supervised}, where the search system balances the ranking information from multiple sources to come up with a ``meta'' search result list that best agree with the individual rankings.
However, these methods have not been applied in the area of faceted browsing to accommodate not only the ranking information but also the users' preference reflected by their facet filter selection actions, which is a primary goal of our work.

\section{Soft Faceted Browsing}

%In this section, we introduce the formalism underlying our soft faceted browsing scheme.
%We first propose a Bayesian probabilistic framework for probabilistically inferring a user's preference given the user's facet selection action, where both the user's preference and action are characterized via probabilistic models.
%Then, we study some representative facets in e-commerce search engines and present a second layer of Bayesian inference framework for learning the user action model in each case.
% \comment{add here a small paragraph to define soft faceted browsing and reiterate its novelty in the context of related work, perhaps by making a comparison with the most similar existing interface work. Then we say that the main technical challenge is to model a user's preference probabilistically, and we propose a Bayesian framework for solving this problem}

%\subsection{Bayesian Probabilistic Framework}

In traditional faceted browsing schemes, when a user selects a particular facet filter, the system shows and only shows the items satisfying the filter, and ranks them in their original order.
Equivalently, the system is implicitly assuming that users always act in a perfectly precise manner in the sense that what users are looking for always satisfy their selected filter.
As is previously mentioned, such a na{\"i}ve assumption barely holds in reality.
Due to the probabilistic nature of human actions, a user may sometimes select a facet filter that does not match the user's interest.
% either due to the absence of a perfectly matching filter or due to the exploratory nature of the user's navigation,
% though the probability of such events happening would generally be low.
A natural idea to characterize such a scenario would be to develop a probabilistic model for users' actions, i.e. a probabilistic distribution over the set of all possible facet filters a user might select given their interest. 

\subsection{A Bayesian Probabilistic Framework}

To formally model the system's task of ranking both within-filter and out-of-filter items in a single ranked list, we postulate that the system should always rank items in descending order of the user's interest in each item as estimated by the system.
Such reasoning is in line with the classic Probability Ranking Principle (PRP) theory \cite{robertson:1977:jdoc} underlying all modern information retrieval systems, which states that a search system should always rank information items in descending order of their probabilities of relevance.
Further, we propose that before the user selects a facet filter (e.g. when the user issues a query to the system), the system has a \emph{prior} estimate of the user's interest in each item (e.g. based on the relevance score of each item with respect to the user's query), and when the user selects a facet filter, the system computes a \emph{posterior} estimate of the user's interest in each item based on the prior estimate and the user's selected facet filter, and the system re-ranks the items in descending order of this posterior estimate.
Intuitively, the system's posterior estimate of the user's interest in an item is dependent upon (a) the prior estimate, and (b) whether the user would be likely to pick the selected facet filter if the user is in fact interested in this item.
In other words, the user's facet filter selection action serves as an additional supporting ``hint'' to the system for probabilistically re-inferring the user's interest.
We now formally cast all such intuitions into a principled Bayesian probabilistic framework:

\begin{definition}[Bayesian ranking principle]
When observing the user's selection of facet filter $a$, the system ranks information items in descending order of the user's \emph{posterior propensity} $p(e|a)$ in each item $e$ that is derived from the user's \emph{prior propensity} $p(e)$ and the user's \emph{action model} $p(a|e)$ via Bayes' theorem:
\begin{equation}
    p(e|a) \propto p(e) \, p(a|e)
\end{equation}
\end{definition}

According to this Bayesian probabilistic framework, the item ranking after the filter selection depends on two components: the prior propensity $p(e)$ and the user action model $p(a|e)$.
The prior propensity characterizes the system's belief \emph{before} the filter selection on how likely the user is interested in each item.
We do not discuss in detail about the prior propensity in this work.
In practice, the prior propensity could be estimated in different ways from the initial ranking as well as additional personalization information of the user if available, and is generally available to us in a search system which ranks items based on probability of relevance.

The action model characterizes how likely the user selects a filter given their interest in a particular item, and is the focus in the rest of this study.
If we restrict $p(a|e)$ so that it equals 1 if and only if $e$ satisfies $a$, and 0 otherwise, then it is easily observed that our Bayesian probabilistic framework is reduced to the traditional faceted browsing scheme, where only the items satisfying the applied filter are returned and ranked in their original order.
On the other hand, if we relax such restrictions and allow $p(a|e)$ to be a true probabilistic distribution, we are essentially capturing the uncertainty underlying user actions and would thus lead to smarter interactions between users and the system.

The Bayesian probabilistic framework could be applied iteratively in cases where the user selects additional facet filters: the posterior propensity would simply serve as the prior propensity for the next user action.
Such a desirable property is a natural consequence of the Bayesian formalism.

The action model in a more general setting can probabilistically characterize a much wider range of user actions in addition to facet filter selections, such as query reformulation, conversational interactions with the system, etc.
The Bayesian probabilistic framework we propose here could serve to provide formal guidance in such scenarios, thus opening up many interesting directions for future research. 

\subsection{Action Model and Inference}

In this section, we discuss concrete ways to instantiate the probabilistic distribution underlying the action model in the cases of some example facets in e-commerce search systems, and we introduce Bayesian inference methodologies for parameter estimation for each probabilistic distribution based on user search log.
The example facets are representative of the major types of facets, and the e-commerce search engine is a typical example of a faceted browsing system.
Thus, the techniques discussed here are universal and could generally be applied to other facets and other faceted browsing systems.

\subsubsection{Brand}

In a typical faceted browsing system, a lot of facets take values from an \emph{unordered} set of values.
One example is the brand facet in an e-commerce search system, where each product has a brand value from the set of possible brand names.
In reality, the brand of the products users are interested in may often be different from the brand they select as filters when they are interacting with the system.
Such phenomena are naturally due to the fact that some brands are similar to each other, so that the users selecting one brand might also be interested in some products of another brand.

The action model in the case of brand, $p(b|e)$, would capture how likely a user would select each brand filter $b$ given the user's interest in a product $e$, and the most natural choice for such an action model is the categorical distribution.\footnote{Since we are characterizing individual brand filter selection actions rather than multiple selection actions as a whole, we use the categorical distribution instead of the multinomial distribution.}
Specifically, suppose there are altogether $k$ brands: $b_{(1)}, b_{(2)}, \ldots, b_{(k)}$.
% then:
% \begin{equation}
%     b|e \sim \mathcal{CAT}(\bm{p}_e)
% \end{equation}
Then for each product $e$, there exists a probability vector $\bm{p}_e = (p_{e(1)}, p_{e(2)}, \ldots, p_{e(k)})$ such that:
\begin{equation}
    p(b_{(i)}|e) = p_{e(i)}\,, \quad i = 1, 2, \ldots k
\end{equation}

Note that if $b_{(i*)}$ is the brand of $e$, then the probability $p_{e(i*)}$ should typically be the highest among all $p_{e(i)}$'s, and the probabilities corresponding to brands similar to $b_{(i*)}$ should generally be higher than those corresponding to more distant brands.
(In the extreme case where we set $p_{e(i*)}$ to be 1 and all other $p_{e(i)}$'s to be zero, it could be easily observed that our model reduces to the traditional ``hard'' filter.)

% To estimate $\bm{p}_e$ for a product $e$, we collect from the user search log the brand filters selected in all search sessions leading to an eventual purchase of $e$.
% Then, we rely on the conjugacy relationship between the Dirichlet distribution and the multinomial distribution for making inference about the vector $\bm{p}_e$ \cite{cowles:2013}.

To estimate $\bm{p}_e$ for a product $e$, we rely on the conjugacy relationship between the Dirichlet distribution and the categorical distribution \cite{cowles:2013}.
In particular, let the prior distribution of $\bm{p}_e$ be:
\begin{equation}
    \bm{p}_e \sim \mathcal{DIR}(\bm{\alpha}_{e}^{(0)})
\end{equation}
where $\mathcal{DIR}(\bm{\alpha}_e^{(0)})$ is a Dirichlet prior with the hyper-parameter vector $\bm{\alpha}_e^{(0)} = (\alpha_{e(1)}^{(0)}, \alpha_{e(2)}^{(0)}, \ldots, \alpha_{e(k)}^{(0)})$.
\footnote{The superscript ``$(0)$'' is used to label the prior hyper-parameters.}
The hyper-parameters $\alpha_{e(i)}^{(0)}$'s, in practice, should be set to reflect the prior belief regarding the degrees to which each brand is related to the product.
For example, in the most na{\"i}ve scenario, if $b_{(i*)}$ is the brand of $e$, then $\alpha_{e(i*)}^{(0)}$ could be a non-zero value and all other $\alpha_{e(i)}^{(0)}$'s are zero.

Next, to make inference, we collect the brand filters selected in any search session leading to an eventual purchase of $e$, and we denote the vector of these brand filters by $\bm{b}_e^{(n)} = (b_1, b_2, \ldots b_n)$.
\footnote{The superscript ``$(n)$'' is used to label the observation vector of size $n$ as well as the posterior hyper-parameters estimated after seeing the observations.}
Then, the posterior distribution of $\bm{p}_e$ could be derived from its prior distribution and observations $\bm{b}_e^{(n)}$ based on Bayes' theorem:
\begin{equation}
    p(\bm{p}_e | \bm{b}_e^{(n)})
    \propto p(\bm{p}_e) \, p(\bm{b}_e^{(n)} | \bm{p}_e)
\end{equation}

Due to the property of conjugacy, the posterior also takes the form of a Dirichlet distribution:
\begin{equation}
    \bm{p}_e | \bm{b}_e^{(n)} \sim \mathcal{DIR}(\bm{\alpha}_{e}^{(n)})
\end{equation}
where $\bm{\alpha}_e^{(n)} = (\alpha_{e(1)}^{(n)}, \alpha_{e(2)}^{(n)}, \ldots, \alpha_{e(k)}^{(n)})$ is the hyper-parameter vector in the Dirichlet posterior that is updated from $\bm{\alpha}_e^{(0)}$ based on:
\begin{equation}
    \alpha_{e(i)}^{(n)} = \alpha_{e(i)}^{(0)} + \sum_{j=1}^{n} \mathds{1}_{\{b_j=b_{(i)}\}}\,, \quad i = 1, 2, \ldots k
\end{equation}
where ``$\mathds{1}$'' is the identity function that takes value 1 if the condition is satisfied and 0 otherwise.
Note that such a posterior update procedure could continue on and on when new observations are obtained from the search log due to the property of conjugacy.

The posterior estimate of $\bm{p}_e$ could either be derived from a \emph{maximum a posteriori} point estimate, or from a posterior predictive distribution.
In the case of the Dirichlet-Categorical conjugacy, these two alternative methods lead to the identical estimates:
% \begin{equation}
%     \hat{p}_{e(i)} = \dfrac{\alpha_{e(i)}^{(n)}}{\sum_{i'=1}^{k} \alpha_{e(i')}^{(n)}}\,, \quad i = 1, 2, \ldots k
% \end{equation}
\begin{equation}
    \hat{p}_{e(i)} = \dfrac{ \alpha_{e(i)}^{(n)} }{  \sum_{i'=1}^{k} \alpha_{e(i')}^{(n)} }\,, \quad i = 1, 2, \ldots k
\end{equation}

\subsubsection{Price}

In contrast to unordered facet value sets, facet values in many cases are \emph{ordinal}, where there is a strict total ordering within the set of all possible facet values.
The categorical distribution in such a scenario is unable to capture the ordinal relationships among the filters.
In an e-commerce search engine, the price facet is one of many examples.
The filters corresponding to the ordinal facets are often presented in \emph{ranges}, e.g. ``\$150-\$200''.
It is observed that when a user selects a particular price range filter, they tend to be more interested in some products priced at around the middle of the range rather than products with prices near the boundary, and there is some slight chance that they might be interested in some products priced completely outside the range, e.g. some product just a little above the range yet having a substantial value relative to its price.

We employ the Gaussian distribution to derive the action model for price facet $p(r|e)$ - the probability of a user interested in a product $e$ selecting a price range $r$.
We postulate that given the user is interested in a product $e$, the probability density of the user selecting a particular price value $c$ follows a Gaussian distribution:
$ c|e \sim \mathcal{N} (\mu_e, \sigma_e^2) $,
and the cumulative probability of the user selecting a particular price range filter $r = [a_r, b_r]$ is computed via integration of the Gaussian density function:
\begin{equation}
    p(r|e) = \Phi\left(\dfrac{b_r-\mu_e}{\sigma_e}\right)
        - \Phi\left(\dfrac{a_r-\mu_e}{\sigma_e}\right)
\end{equation}
where $\Phi$ denotes the cumulative distribution function of the standard Gaussian distribution.
Note that $\mu_e$ should typically be close to the price of $e$, $c_e$, but users may not have a precise idea of the price of their interested product, so we treat $\mu_e$ as unknown and learn its value from user search log.
Under such a Gaussian model for price filter selection actions, given that a user is interested in a particular product with a price tag $c$, they may most likely select a price range filter that covers $c$ at around the mid-point of the range, and would less likely select a price range filter with its boundary very close to $c$, and it would be even less likely but not impossible that they select some price range filter not covering $c$ at all.
Such consequences nicely coincide with our intuition.
(In the extreme case where we set $\mu_e = c_e$ and $\sigma_e^2 \to 0$, it could be easily observed that our model reduces to the traditional ``hard'' filter.)

% Both $\mu_e$ and $\sigma_e^2$ in the Gaussian model are unknown and need to be estimated.
% Again, we collect from user search log the price range filters selected in all search sessions leading to an eventual purchase of $e$.
% To make the inference computation tractable, we pick the mid-point of each filter to approximate the entire range for making the inference.
% Then, we deploy the conjugacy relationship between the Normal-Inverse-Gamma distribution and the Gaussian distribution to estimate $\mu_e$ and $\sigma_e^2$ \cite{cowles:2013}.

The two parameters $\mu_e$ and $\sigma_e^2$ are typically unknown in the real world, so we need to make inference from observations of past user activities.
% We introduce the Bayesian inference procedures for the utility action model as a principled method to instantiate the SPF model in practice based on search log.
In Bayesian statistics theories, the Gaussian distribution with both its mean and variance unknown has the Normal-Inverse-Gamma ($\mathcal{NIG}$) distribution as its conjugate prior.
Thus, we define the prior distribution for $\mu_e$ and $\sigma_e^2$ as:
% \footnote{The superscript ``$(0)$'' is used to label the prior hyper-parameters.}
% \begin{definition}[UAM prior distribution]
\begin{equation}
    \mu_e, \sigma_e^2 \sim \mathcal{NIG}(\bm{\mathcal{H}}_{e}^{(0)})
\end{equation}
$\bm{\mathcal{H}}_{e}^{(0)} = (\mu_{e}^{(0)}, \kappa_{e}^{(0)}, \alpha_{e}^{(0)}, \beta_{e}^{(0)})$ represents the hyper-parameter vector in the $\mathcal{NIG}$ prior in the form of:
\begin{empheq}[left=\empheqlbrace]{align}
    \; \sigma_e^2 | \alpha_{e}^{(0)}, \beta_{e}^{(0)} & \sim \mathcal{IG}(\alpha_{e}^{(0)}, \beta_{e}^{(0)}) \\
    \; \mu_e | \sigma_e^2, \mu_{e}^{(0)}, \kappa_{e}^{(0)} & \sim \mathcal{N}(\mu_{e}^{(0)}, \sigma_e^2 / \kappa_{e}^{(0)})
\end{empheq}
% \end{definition}
where ``$\mathcal{IG}$'' denotes the inverse gamma distribution.
In practice, the hyper-parameters could be heuristically set based on any available prior knowledge about how users select price ranges.
For example, we typically set $\mu_{e}^{(0)} = c_e$.

% In the search log of a contemporary e-commerce search system, the users' eventual purchases (if any) together with their selected price filters (if any) are typically recorded for each search session.
% Therefore, for each product $e$, the price filters selected in all the sessions that resulted in an eventual purchase of $e$ could be collected to form the set of observations for making inference on the utility of $e$.
For each product $e$, we collect the price filters selected in all the sessions that resulted in an eventual purchase of $e$ from the search log and form the set of observations for making inference on $\mu_e$ and $\sigma_e^2$.
To make the inference computation tractable, we pick the mid point $m_r$ of the price range in each selected filter $r$ as an approximation to the whole range. \footnote{Since the price ranges of the price filters in e-commerce search engines are often relatively short segments as compared to the magnitude of product prices, it is typically reasonable to approximate the whole ranges by their mid points.}
Thus, for each product $e$, we obtain an observation vector composed of the mid points of all the selected price filters for product $e$, and we denote the vector by $\bm{m}_e^{(n)} = (m_1, m_2, \ldots m_n)$.
% \footnote{The superscript ``$(n)$'' is used to label the observation vector of size $n$ as well as the posterior hyper-parameters estimated after the observation.}
Following that, the posterior distribution for $\mu_e$ and $\sigma_e^2$ could be derived from their prior distribution and observations $\bm{m}_e^{(n)}$ based on Bayes' theorem:
\begin{equation}
    p(\mu_e, \sigma_e^2 | \bm{m}_e^{(n)})
    \propto p(\mu_e, \sigma_e^2) \, p(\bm{m}_e^{(n)} | \mu_e, \sigma_e^2)
\end{equation}

Due to the property of conjugacy, the posterior also takes the form of an $\mathcal{NIG}$ distribution:
% \begin{definition}[UAM posterior distribution]
\begin{equation}
    \mu_e, \sigma_e^2 | \bm{m}_e^{(n)} \sim \mathcal{NIG}(\bm{\mathcal{H}}_{e}^{(n)})
\end{equation}
where $\bm{\mathcal{H}}_{e}^{(n)} = (\mu_{e}^{(n)}, \kappa_{e}^{(n)}, \alpha_{e}^{(n)}, \beta_{e}^{(n)})$ is the hyper-parameter vector in the $\mathcal{NIG}$ posterior that is updated from $\bm{\mathcal{H}}_{e}^{(0)}$ based on the sample mean $\overline{m}_e$ and variance $s_e^2$ of $\bm{m}_e^{(n)}$:
% \begin{empheq}[left=\empheqlbrace]{align}
    % \; \mu_{e}^{(n)} = & \left.\left( \kappa_{e}^{(0)} \mu_{e}^{(0)} + n\,\overline{m}_e \right) \middle/ \left( \kappa_{e}^{(0)} + n \right)\right. \\
    % \; \kappa_{e}^{(n)} = & \ \kappa_{e}^{(0)} + n \\
    % \; \alpha_{e}^{(n)} = & \ \alpha_{e}^{(0)} + n/2 \\
    % \; \beta_{e}^{(n)} = & \ \beta_{e}^{(0)} + \left.(n-1)s_e^2 \middle/ 2\right. \\
    % & + \left.\kappa_{e}^{(0)} n \left( \mu_{e}^{(0)} - \overline{m}_e \right)^2 \middle/ \left(2\kappa_{e}^{(0)} + 2n\right)\right.
% \end{empheq}
\begin{empheq}[left=\empheqlbrace]{align}
    \; \mu_{e}^{(n)} = & \dfrac{ \kappa_{e}^{(0)} \mu_{e}^{(0)} + n\,\overline{m}_e }{ \kappa_{e}^{(0)} + n} \\
    \; \kappa_{e}^{(n)} = & \ \kappa_{e}^{(0)} + n \\
    \; \alpha_{e}^{(n)} = & \ \alpha_{e}^{(0)} + \dfrac{n}{2} \\
    \; \beta_{e}^{(n)} = & \ \beta_{e}^{(0)} + \dfrac{(n-1)s_e^2}{2}
    + \dfrac{\kappa_{e}^{(0)} n \left( \mu_{e}^{(0)} - \overline{m}_e \right)^2}{\left(2\kappa_{e}^{(0)} + 2n\right)}
\end{empheq}
% \end{definition}

Note again that such a posterior update procedure could continue on and on when new observations are obtained from the search log, due to the property of conjugacy.

With the posterior distribution of $\mu_e$ and $\sigma_e^2$ established, two alternative methods could be used for estimating the Gaussian model for price filter selection actions.
The first one is directly based on the \emph{maximum a posteriori} point estimate from the posterior:
% \begin{definition}[Simple utility action model]
\begin{equation}
    \hat{p}(r|e) = \Phi\left(\dfrac{b_r-\hat{\mu}_e}{\hat{\sigma}_e}\right)
            - \Phi\left(\dfrac{a_r-\hat{\mu}_e}{\hat{\sigma}_e}\right)
            % \Phi(\dfrac{b_r-\hat{\mu}_e}{\hat{\sigma}_e}) - \Phi(\dfrac{a_r-\hat{\mu}_e}{\hat{\sigma}_e})
    % = \mathfrak{I}(\hat{\mu}_e, \hat{\sigma}_e^2, a_r, b_r)
\end{equation}
where $\hat{\mu}_e$ and $\hat{\sigma}_e^2$ could be shown to come from:
\begin{empheq}[left=\empheqlbrace]{align}
    \; \hat{\mu}_e & = \mu_{e}^{(n)} \\
    \; \hat{\sigma}_e^2 & = \dfrac{\beta_{e}^{(n)}}{\alpha_{e}^{(n)} + \dfrac{3}{2}}.
\end{empheq}
% \end{definition}

The second estimation method comes from the posterior predictive distribution, which, in the case of Normal-Inverse-Gamma distribution, could be shown to follow a $t$-distribution:
% \begin{definition}[Posterior predictive utility action model]
\begin{equation}
    % \tilde{p}(r|e) = \int_{\sigma_e^2} \int_{\mu_e} \mathfrak{I}(\mu_e, \sigma_e^2, a_r, b_r) \, p(\mu_e, \sigma_e^2 | \bm{m}_e) \, d\mu_e \, d\sigma_e^2
    \tilde{p}(r|e) = \Psi \left( \dfrac{ b_r - \tilde{\mu}_{e} }{ \tilde{\sigma}_e }\right)
    - \Psi \left( \dfrac{ a_r - \tilde{\mu}_{e} }{ \tilde{\sigma}_e }\right)
    % \tilde{p}(r|e) = \Psi_{ 2\alpha_{e}^{(n)} } \left(
    %         \left.\left( b_r - \mu_{e}^{(n)} \right)\middle/ \tilde{\sigma}_e^{(n)} \right.
    % \right) - \Psi_{ 2\alpha_{e}^{(n)} } \left(
    %         \left.\left( a_r - \mu_{e}^{(n)} \right)\middle/ \tilde{\sigma}_e^{(n)} \right.
    % \right)
    % \tilde{p}(r|e) = \Psi_{ 2\alpha_{e}^{(n)} } \left(
    %         \left( b_r - \mu_{e}^{(n)} \right) \middle/ \sqrt{ \dfrac{ \beta_{e}^{(n)} \left(\kappa_{e}^{(n)}+1\right) }{ \alpha_{e}^{(n)} \kappa_{e}^{(n)} } }
    % \right) - \Psi_{ 2\alpha_{e}^{(n)} } \left(
    %         \left.\left( a_r - \mu_{e}^{(n)} \right)\middle/ \sqrt{ \dfrac{ \beta_{e}^{(n)} \left(\kappa_{e}^{(n)}+1\right) }{ \alpha_{e}^{(n)} \kappa_{e}^{(n)} } } \right.
    % \right)
\end{equation}
where $\Psi$ denotes the cumulative distribution function of a $t$-distribution with $2\alpha_{e}^{(n)}$ degrees of freedom, and $\tilde{\mu}_{e}$, $ \tilde{\sigma}_e^2 $ come from:
\begin{empheq}[left=\empheqlbrace]{align}
    \; \tilde{\mu}_{e} & = \mu_{e}^{(n)} \\
    \; \tilde{\sigma}_e^2 & = \dfrac{ \beta_{e}^{(n)} \left(\kappa_{e}^{(n)}+1\right) }{ \alpha_{e}^{(n)} \kappa_{e}^{(n)} }
\end{empheq}
% \end{definition}

In practice, these two methods lead to almost identical inference results given a moderately large observation vector.
In our experiments, we always employ the first method.

\section{Experiments}

To demonstrate the effectiveness of our proposed soft faceted browsing scheme, we implemented our Bayesian probabilistic framework in an internal prototype system on top of the Walmart e-commerce search engine \cite{walmart}.
% Figure 1 shows the outcomes of applying the traditional faceted browsing scheme (the middle figure) and our new soft faceted browsing scheme (the right figure) on a ranked list for the query ``luggage'' (the left figure), as a response to the user's selection of the ``[\$20, \$50]'' filter.
% We see that two additional items were inserted into the list that are outside the user's selected price range, and they turn out to be decent choices for the user though they're priced slightly above \$50.
We performed extensive experiments using search log data to compare the soft faceted browsing scheme against the traditional faceted browsing scheme when applied to the price and brand filters.
The experiments for the price and brand filters were very similar; we only outline the experiment set-up and results for the price filters.

We collected around 62,000 search sessions in a recent month in which the user (a) selected at least one price filter (which was implemented in the ``hard'' filter manner) and (b) purchased one product at the end of the session.
Without doing any advanced computation, we already noticed clear evidence in support of our motivating intuitions regarding the weaknesses of ``hard'' filters: around 43\% of the users' selected price filters did \emph{not} cover the price of their eventually purchased product, and in such cases the user had to de-select the filter and either scanned the unfiltered list or tried other price filters and/or other facet selections so as to navigate to the product of their interest, which greatly affected the navigational efficiency as well as the overall user experience.

Then, we carried out simulated user evaluations relying on this search log dataset.
We collected 20 most popular queries that led to at least 700 purchases in the month,
% Most of these queries are short, single word queries, e.g. ``tvs'', ''phones'' etc., while a few of them are longer with the users' desired attributes of the products, e.g. ``straight talk phones'', ``tvs on clearance'', etc.
and we performed leave-one-out cross validations to compare our proposed soft faceted browsing scheme with the traditional ``hard'' faceted browsing scheme via simulated user interactions with the system.
For each query, among all filters ever selected by any user who issued the query and eventually made a purchase, we treated one of them as the test data and all the rest as the training data at a time.
Specifically, we trained our model using all but one of the filters to learn the user action model for all the products returned by the search engine, and then applied the trained model to the filter that was left as the test data and recorded down the rank of the user's eventually purchased product.
Meanwhile, we also applied the traditional ``hard'' filter to all the products returned by the search engine, and recorded the rank of the user's eventually purchased item.
In cases where the ``hard'' filter missed the user's eventually purchased item, we tried not to ``over-penalize'' the ``hard'' filter: we computed the rank as the total number of filtered products plus the rank of the user's purchased product in the unfiltered list, emulating the scenario where the user scans the entire filtered list without finding the product, de-selects the filter, and then scans the unfiltered list to look for the product.
We performed one-sided Wilcoxon signed-rank tests for the comparison for each of the 20 queries.
% and we heuristically set the prior parameters for our model to minimize the average of the 20 p-values.
We observe that 19 out of the 20 queries have a p-value less than $10^{-6}$, strongly indicating that our soft faceted browsing scheme is significantly superior than the traditional ``hard'' faceted browsing scheme in terms of its efficiency in helping users navigate to the products of their interests.
The only one exception is the query ``electronics'' with p-value $0.00187$ (which is also very significant though not as extreme as the other queries), and the reason is observed to be that most of the users' purchases are both (a) covered by the filters they selected and (b) concentrated on the very top portion of the filtered list, in which cases the ``hard'' filter is relatively harder to beat.

\section{Conclusions and Future Work}

We proposed a novel soft faceted browsing scheme for information access systems where, when the user selects a facet filter, the system may return a few relevant items that do not satisfy the filter in a non-intrusive way alongside the items that satisfy the filter.
We provided a formal Bayesian probabilistic framework for realizing such a soft faceted browsing scheme that takes into consideration the probabilistic nature of users' facet filter selection actions, and we demonstrated that our method is more effective than traditional ``hard'' faceted browsing scheme via experiments using e-commerce search log data.
We mainly relied on simulated user studies in this work, and we will be conducting real user experiments using the prototype systems we built for further comparison studies as the next step.

The proposed framework and model open up interesting new research opportunities in the intersection of machine learning and information retrieval.
An interesting extension is to introduce active learning for optimal preference elicitation (e.g., dynamically adjust the price ranges to focus on the most uncertain range of prices). 

\bibliographystyle{plain}
\bibliography{main}

\end{document}